# Fault Detection using Immune-Based Systems and Formal Language Algorithms

*J.F. Martins\*[#], P. J. Costa Branco[#], A.J. Pires\*[#] and J.A. Dente[#]*

[#]*Laboratório de Mecatrónica*
*Instituto Superior Técnico (I.S.T.)*
*Av. Rovisco Pais, 1096 Lisboa Codex*
*E-mail: pbranco@alfa.ist.utl.pt*
*http://pbranco.ist.utl.pt*
*Portugal*

[\*]*Escola Superior de Tecnologia de Setúbal*
*Instituto Politécnico de Setúbal*
*Rua do Vale de Chaves, Estefanilha 2910 Setúbal*
*Tel: 351-265790000 Fax: 351-265721869*
*E-mail: jmartins@est.ips.pt*
*Portugal*

Abstract: This paper describes two approaches for fault detection: an immune-based mechanism and a formal language algorithm. The first one is based on the feature of immune systems in distinguish any foreign cell from the body's own cell. The formal language approach assumes the system as a linguistic source capable of generating a certain language, characterised by a grammar. Each algorithm has particular characteristics, which are analysed in the paper, namely in what cases they can be used with advantage. To test their practicality, both approaches were applied on the problem of fault detection in an induction motor.

## I. INTRODUCTION

Classification algorithms have been extensively researched for fault diagnosis. These techniques, however, demand prior knowledge about particular system malfunctions since they search for specific fault patterns. The detection system becomes then vulnerable when dealing with different faults. The use of model-based fault detection has represented another approach. However, when the relationship between the system variables are difficult to establish and/or the data available is limited, the extracted system model can fail in giving enough accuracy responses to distinguish a fault from a non-fault situation. Novelty-detection algorithms, when compared with these approaches, have significant differences. They detect probabilistically any unacceptable abnormality rather than looking for specific patterns. Allowing to early detecting an abnormal situation, providing the fault not being immediately disastrous, allows the use of preventive measures to avoid any subsequent damage.

In this paper, two fault detection algorithms are presented. The first one, a novelty-detection algorithm inspired in immune systems, was based on [5] and provides better characteristics to engineering applications. The second algorithm, based on formal language theory [1,3], was developed for dynamical systems applications. Based on their analysis and discussion, preliminary results are reported, illustrating their capability to deal with the problem of monitoring electric machines.

## II. IMMUNE-BASED FAULT DETECTION SYSTEM

This approach uses the property of immune systems in distinguish any foreign cell (*non-self*) from the body's own cell (*self*) [4]. When transposed to fault detection, this characteristic can be used to discriminate normal system patterns from any non-tolerable pattern deviations, resulting in a novelty-detection algorithm. It is important to identify some cases where this characteristic could represent an advantage:

(a) When the normal situation of a system is characterised by a set of complicated patterns, it is difficult to extract their relations. In this case, it is interesting not to ponder the normal patterns but its complement, the abnormal situations.

(b) There are systems where the possible abnormal patterns can consist in a much larger number than the normal ones. Since to train a fault detection system with a large number of fault situations becomes unpractical, it is preferable to first detect any abnormal situation and after effectuate their classification as some detectors are activated.

An algorithm proposed by Dasgupta [5] and also based on the censoring mechanism occurring with T-cells in the thymus has been proposed to fault detection. Since there is discrimination between normal from abnormal patterns, fault detection is based on distance and similarity measures implemented by a fixed matching rule, where a pattern is compared with the ones in a detector set.

The matching rule used by Dasgupta considers a match between two binary strings in a bit-a-bit basis. Two string patterns match if their bits agree at least $r$ contiguous. This matching rule had been adequate for the problem of computer virus detection, as demonstrated in [5], since changes can be made at a bit level. However, a matching rule of this type is not suitable for the engineering problems of fault detection since it looses any physical meaning related with the process to be monitored. Despite this, the matching rule considers all bits as having equal significance, which makes the detectors too sensitive to noise effects, detecting false faults. Therefore, a more suitable matching rule is proposed in this paper, which measures the similarity between two patterns by using the Euclidean distance.

*II.1 The proposed immune-based algorithm.*

Figure 1 presents a schematic diagram of the algorithm with its five steps:

(1) *Process data acquisition.* Establish a set of data representative of the normal activity of the system to be monitored;

(2) *Set the parameters values.* Set the number of bits ($n$), the data window size ($w$), and the number of detectors to be generated ($d$). In the matching rule, a match occurs between two patterns when their Euclidean distance is less than a percentage ($md$) of the highest distance in the hypercube that forms the patterns domain.

(3) *Encoding*. Normalise the data, follow to the binary codification, and perform the window tracking of data as shown in figure 1. This produces the data set $S$ representing the *self*.

(4) *Detectors generation.* Randomly generate a possible detector. Verify the matching distance between the detector and each pattern of $S$, and also with to the detectors already generated. If some distance is smaller than the matching value, reject this detector and begin this step to generate another one. On other hand, if all distances computed are superior to the matching value, this is a valid detector. This censoring procedure (*negative selection*) duplicates the process occurred in our thymus during the first years after birth with the lymphocyte cells (*T-cells*). At last, verify if the number of valid detectors is less than $d$. If yes, begin step again.

This step introduces an important advantage to our method. In the original algorithm, the detectors generated are not removed or checked for possible duplications. Therefore, one cannot guaranty that the set of detectors cover the patterns domain enough. Some detectors could be so near that their representative patterns can be regarded as the same, which effectively decreases the monitoring performance. When considering a minimum matching distance, the detectors become better distributed by the domain, which reduces the possibility of concentration of detectors in a region of the domain.

(5) *Monitoring process.* Once the detectors had been produced, the condition of the system can be continually monitored matching its patterns against the detectors. The incoming pattern is compared with all detectors. If any distance value is smaller than the matching value, the respective detector is activated and thus an abnormal situation is known to have occurred.

Suppose a simple monitoring case where a sinusoidal signal $y(t) = \sin(t)$ is our normal pattern to be monitored (*self*). Two abnormal situations were created. In the first one, the signal had its frequency increased by 10%. In the second abnormal situation, a more severe distortion was introduced, which resulted in a signal as $y(t) = 0.5\sin(t) + 0.5\sin(2t)$.

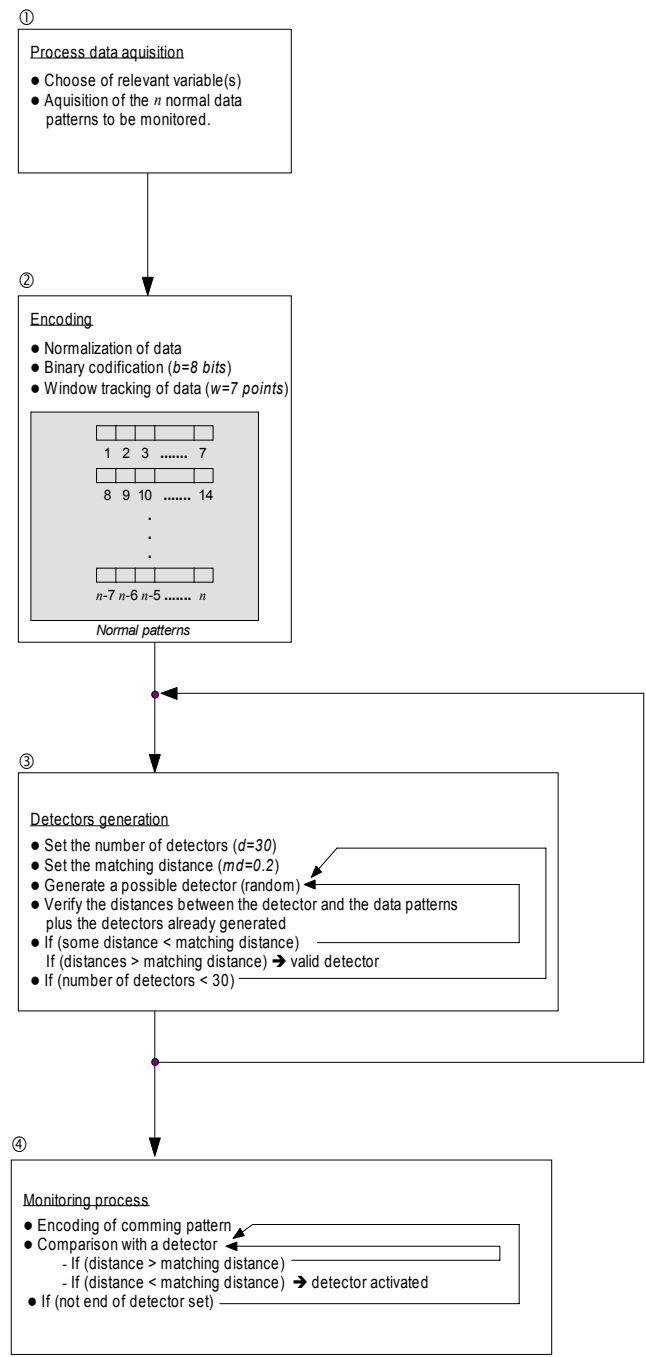

*Fig. 1 Diagram of the new novelty-detection algorithm*

Figure 2(a) shows the sinusoidal signal that represents the normal condition (*self*) by a dashed line. The two abnormal situations are shown in solid line. The parameters used in the algorithm were $b=8$, $w=7$, and $md=0.2$. A set of 30 detectors were generated and used to monitoring. Figure 2(b) shows the monitoring results. The detectors have been activated regularly, which was caused by the periodicity of the new patterns. Also note that the frequency that the detectors are activated increases, thus indicating the advancing distortion of the signal.

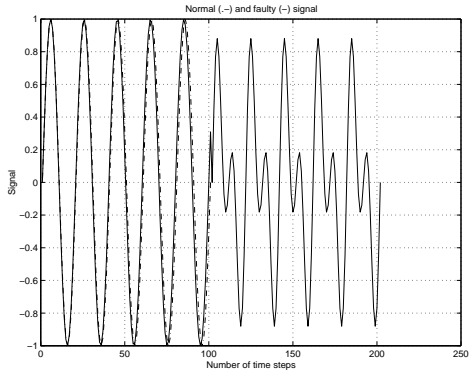

(a)

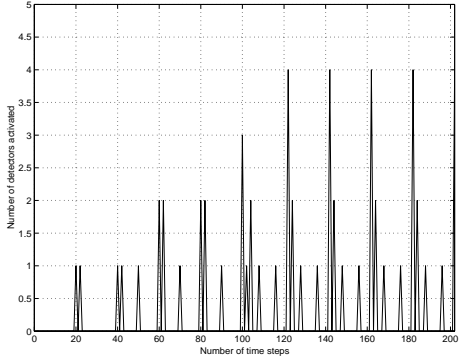

(b)

*Fig. 2 (a) Normal pattern (.-) and abnormal situations (-). (b) Detectors activated by new patterns.*

### III. FORMAL LANGUAGE FAULT DETECTION SYSTEM

The formal language procedure considers a linguistic source that generates the words within the produced language. In this way a dynamical system is assumed as a linguistic source capable of generating a certain language, characterised by a grammar that can generate all the words in that particular language. This grammar, denoted by *G*, defines the structural features of the words produced by the linguistic source and, in this way, models the source itself [1]. *G* is a 4-tuple (1), where the terminal alphabet $\Sigma_T$ is constituted by symbols that make up the resulting words, the non terminal alphabet $\Sigma_N$ is constituted by symbols that are used to generate the patterns, the start symbol S is a special non-terminal symbol that is used to begin the generation of words, and the set of productions P is a set of rules (in the form α→β, where α and β are strings) that determines the generation of words [2].

$$G = (P, \Sigma_N, \Sigma_T, S) \quad (1)$$

Both alphabets must be established from the codification of the variables obtained from the dynamical system. Within the established formal language formalism [6] the codification of the output variable – *y* – produces the terminal alphabet and codification of the input variable – *u* – produces the non-terminal (2).

$$y \xrightarrow{Codification} y_j \in \Sigma_T$$
$$u \xrightarrow{Codification} U_i \in \Sigma_N \quad (2)$$

The learning algorithm must establish the relations between both alphabets (input and output information) in order to produce terminal words that denote the output variable evolution. In this way *p-type productions* are considered assuming the general form (3). Sequence $y_1...y_p$ is constituted by terminal symbols and has of length p, $U_k$ is any non terminal symbol, $y_{p+1}$ is a terminal symbol, and δ is a special non terminal symbol that can be replaced by any non terminal symbol. A *p-type production* has p terminal symbols in the left part of the production, denoting the previous dynamic evolution of the output variable.

$$y_1 \ldots y_p U_k \rightarrow y_1 \ldots y_p y_{p+1} \delta \quad (3)$$

Given this general form production (3), since the left part of any production contains at least one non-terminal symbol, this grammar can be classified as context sensitive in the Chomsky hierarchy [2].

To establish the grammar productions a grammatical inference based algorithm was developed. Out of a set of sample words, resulting from the codification of the dynamic evolution of the system input/output variables, the basic learning algorithm has a simple paradigm in order to establish the set of productions:

- A 0-type production is considered for every symbol of the terminal alphabet that occurs in the sample.
- A (n+1)-type production is considered if the established n-type production already exists.

One must note that the structure of the formal language productions is not established in advance. According to the word samples involved, one can obtain different types of productions (0-type, 1-type, ...) in the resulting grammar. This feature allows the modelling of different behaviours detected in a dynamical system.

The obtained general grammar - *G* - defines a class of patterns represented by strings belonging to the language which the grammar represents - *L(G)*. In this context it is possible to use this grammar to recognise the well formed strings and reject any strings that are in any way imperfect. This is the subjacent idea to fault detection, since a faulty dynamical system will certainly generate strings that do not belong to its initial generated language.

The detection of a fault in a dynamical system is then based on distance and similarity measures, where an unknown string is compared with the ones produced by the proper grammar. The basic algorithm computes the distance between the string generated by the dynamical system and the one generated by the inferred grammar. If this distance exceeds a certain threshold a fault is reported, otherwise the system is considered to work properly.

Several methods for structural word matching have been reported in the literature. The basic algorithm states that the distance between two words is related to the sequence of edit operations (substitution, insertion, and deletion) required to transform one word into another. For any sequence of edit operations a cost function (4) is considered, where *c* denotes the cost of a particular sequence *s*, and *c(e_i)* the cost of a particular edit operation.

$$c(s) = \sum_{i=1}^{n} c(e_i) \quad (4)$$

The distance between two words $w_1$ and $w_2$ is defined as the minimum cost of transforming a word into another (5).

$$d(w_1, w_2) = \min \left\{ c(s) \middle| \begin{array}{l} s \text{ being a sequence of edit} \\ \text{operations transforming} \\ w_1 \text{ into } w_2 \end{array} \right\} \quad (5)$$

Let us assume that a dynamical system is represented by a grammar $G_1$, inferred from random data. If, from certain instant on, the system shall present a fault one can consider that a new grammar $G_2$ should be considered for the same system. Since both grammars are different there must be strings that do not match. For the same sequence of non-terminal symbols and productions the corresponding terminal word should be different, that is, the distance between both words is greater than a certain threshold.

## IV. RESULTS

As an application example, we have considered the use of both algorithms in fault detection of an induction motor. Namely, we considered an experiment in the detection of a rotor cage fault caused by a possible broken bar in the rotor. Due to a broken bar, and consequently variation of the rotor impedance, the rotor m.m.f. will be subject to modulation. A stator winding e.m.f. will be induced and hence a current, which is a measure of rotor asymmetry.

To prepare a normal data set, the motor was first operated without fault. Three different and increasing normal loads were applied to the machine resulting in an increasing phase current, as shown in figure 3.

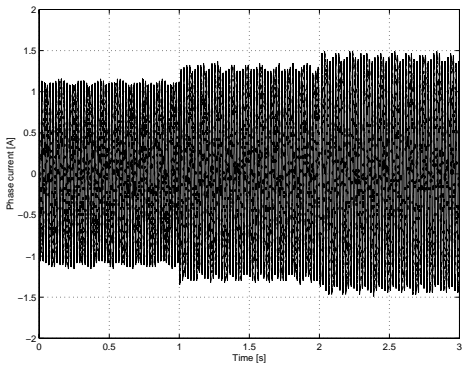

*Fig. 3 Current signal of normal load machine behavior.*

To detect abnormal operation of the motor two tests were considered. The first one, presented in figure 4, considers two distinct functioning conditions. The motor was first operated without load and without fault, followed by loaded broken bar fault condition. Both working conditions weren't considered in the normal data set, testing the generalisation capability of the algorithms.

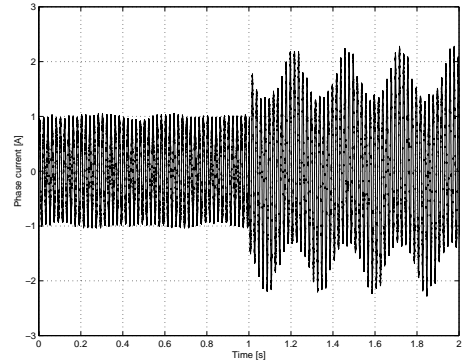

*Fig. 4 Current signal for normal unloaded behaviour follwed by loaded faulty behaviour*

The second test, presented in figure 5, considers a broken bar condition, with four increasing stages of load applied to the induction machine, starting from a no load situation

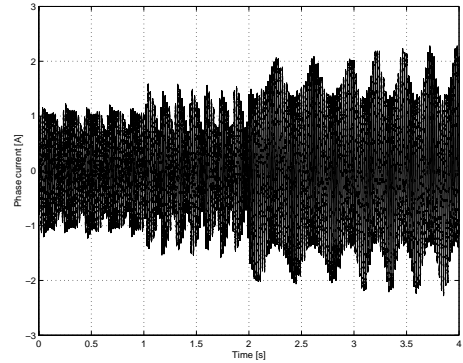

*Fig. 5 Current signal for increasing load abnormal behavior.*

### IV.1. Immune-Based System

To test the algorithm, quantification of the absolute value of one of the stator currents establishes the system patterns. The set *S* of self was constructed using the data shown in Figure 3 that shows the normal evolution of the stator current when operating the motor without fault and three load cases. Based on this data, 30 detectors were generated using as algorithm's parameters: *b=8* (encoding using eight bits), *w=7* (seven data values per window), and *md=0.2*. These were used to monitoring the data shown in figure 4. Note that this data was not used to generate the detectors and so the monitoring process has to demonstrate its generalization capability. In the first part of data of Figure 4, the motor operates in a normal condition but without load. Therefore, we expect that no detector be activated, although they had been generated using only normal load data. In the second part of data, the motor is in a fault situation and with load. However, this load has a higher value than the ones presented in figure 3. Once again, the monitoring process has to show its generalization capability to detect this fault situation.

Figure 6(a) shows the detectors activated during the monitoring process. During normal operation, none detector has been activated. As soon as the system receives abnormal patterns, some detectors are activated revealing an abnormal machine operation. In figure 6(b), the frequency that each detector was activated during the abnormal situation is shown. This information could be used as an indicator of which anomaly occurred, acting as a classification procedure.

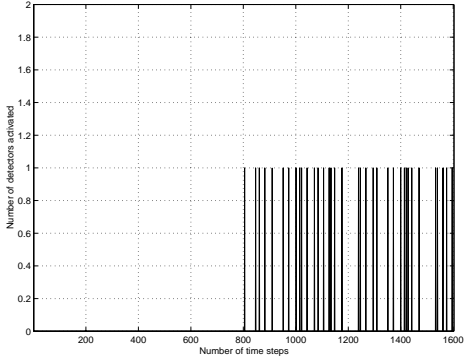

(a)

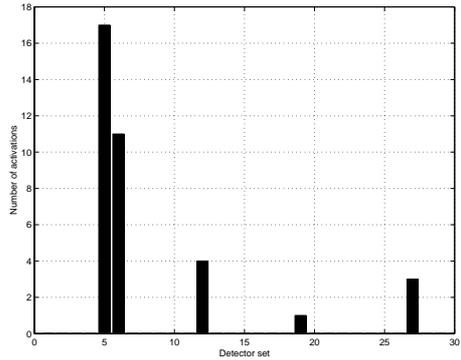

(b)

*Fig. 6 (a) Detectors activated during monitoring process.*
*(b) Histogram for the frequency that each detector was activated.*

Another test was conducted that considers four different subsets of the faulty condition but for increasing loads. In this test, the system generates again 30 detectors from the complete normal set. After generating the detectors figure 7 shows their activation. The results show that for higher applied load a higher number of detectors were activated, since the abnormal patterns become more relevant. Note that to the initial fault situation without load, the number of detectors activated were small. This shows a significant property of the novelty-detection algorithm which response magnitude is proportional to the extent of detected abnormal situation.

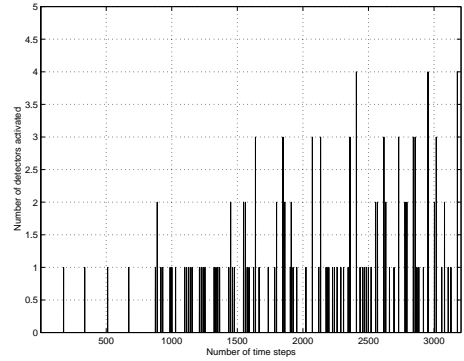

*Fig. 7 Detectors activated during monitoring process.*

*IV.2. Formal Language System*

The inferred grammar represents the evolution of the induction motor without fault, and it is obtained from sample words – figure 2 – of the motor normal operation. The quantification of the stator currents will establish the terminal alphabet. In order to have a better quantification of these currents their are represented in a dq rotating frame model [6] with speed $\omega_R$ (6), so that, this quantification process only depends on the current amplitude. $\omega$ denotes the rotor speed, M the mutual (stator-rotor) induction coefficient, $i_{qs}$ the q-stator current component, $\tau_r$ the rotor time constant, and $\psi_r$ the rotor flux.

$$\omega_R = \omega + \frac{M}{\tau_r} \frac{i_{qs}}{\psi_r} \qquad (6)$$

As a first test we consider the data set already presented before in figure 3, where a loaded faulty situation follows an unloaded non-faulty one. It can be seen from figure 8. that the formal language based fault detection algorithm detects the deviation from the initial grammar, whenever the fault begins. A fault threshold (distance between words) of 10 was considered. A non-zero distance obtained for the unloaded non-faulty operation appears since this condition was not present in the learning phase. However the small distances obtained reveal good generalization properties.

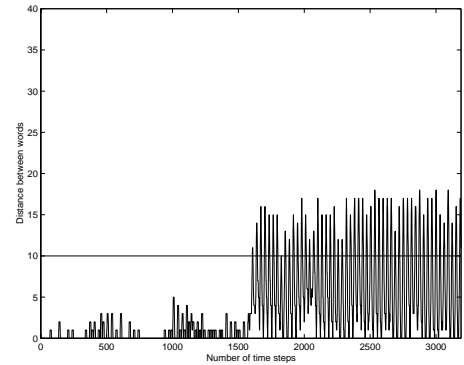

*Fig. 8 Distance between the words generated by the grammar and the induction motor*

The second test, considered before in figure 5, assumes four different subsets of faulty operation. Figure 9 shows that the distance between words is repeatedly above the fault threshold, denoting the presence of a fault. One can verify that for higher loads the signal denoting the word distance presents a higher frequency.

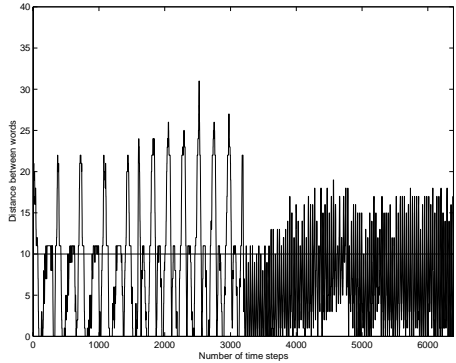

*Fig. 9 Distance between the words generated by the grammar and the induction motor*

V. Discussion, Conclusion and Future Work

In this paper, two new approaches to systems fault detection were presented. Although the immune-based algorithm proposed in [4] was designed to be able to detect computer viruses and adapted in [5] for fault systems detection, its matching rule is inappropriate for fault detection in engineering systems, as it does not guaranty a significant homogeneous coverage of the pattern space. Another matching rule and an improved detectors generation process was proposed and tested with good results. The influence of selected parameters to the algorithm's performance needs further examination. An adaptation mechanism to the detectors is currently being designed. This mechanism allows to move the detectors to other areas of the pattern domain as new data comes from the system to complete the initial data.

The second algorithm performs the fault detection by formal language techniques. Unlike other approaches, the nature of the productions that define the dynamical system relationships are not set up in advance, different types of productions are established on-line according to the incoming words from the linguist source. The fault detection is based on the distance between the words generated by the faulty system and the grammar that represents the non-faulty system. The algorithm presents good generalisation capabilities, however the fault threshold must be accurated chosen. Work is being done in order to relate the word distance, and thus grammar distance, with the type and severity of the fault.